\documentclass[aps,twocolumn,prd]{revtex4}

\def\be{\begin{equation}}
\def\ee{\end{equation}}
\def\bea{\begin{eqnarray}}
\def\eea{\end{eqnarray}}
\def\ba{\begin{array}}
\def\ea{\end{array}}
\def\bc{\begin{center}}
\def\ec{\end{center}}

\def\ghost#1{}

\def\simge{\mathrel{%
   \rlap{\raise 0.511ex \hbox{$>$}}{\lower 0.511ex \hbox{$\sim$}}}}
\def\simle{\mathrel{
   \rlap{\raise 0.511ex \hbox{$<$}}{\lower 0.511ex \hbox{$\sim$}}}}

\def\NPB{{\em Nucl. Phys.} B}
\def\PLB{{\em Phys. Lett.} B}
\def\PRL{{\em Phys. Rev. Lett.}\,\,}
\def\PRD{{\em Phys. Rev.} D}

\begin{document}

\title

\title{CONSTRAINTS \\ [4mm] ON ~THE ~PARITY-VIOLATING ~COUPLINGS \\ [4mm]
OF A ~NEW ~GAUGE ~BOSON\\ [10mm]}

\author{\large C. Bouchiat \ and \ P. Fayet\,$^1$
\vspace{6mm}
\\
$^1$ {\small Laboratoire de Physique Th\'eorique de l'ENS
{\small \ (UMR 8549 CNRS)}} \\
{\small 24 rue Lhomond, 75231 Paris Cedex 05, France}\medskip \\ [-3mm]\ }

\date{october 18, 2004}

\vspace*{3mm}

\begin{abstract}
\ \\ [3mm]
High-energy particle physics experiments allow for the possible existence
of a new light, very weakly coupled, neutral gauge boson (the $U$ boson).
This one permits for {\it \,light\,} (spin-$\frac{1}{2}$ or spin-0) particles to be acceptable Dark Matter candidates,
by inducing sufficient (stronger than weak)
annihilation cross sections into $e^+e^-$.
They could be responsible for the bright \,511 keV $\gamma$ ray line
observed by INTEGRAL from the galactic bulge.
\\[-2mm]\\
Such a new interaction may have important consequences, especially
at lower energies. Parity-violation atomic-physics experiments provide
strong constraints on such a $U$ boson,
if its couplings to quarks and electrons violate parity.
With the constraints coming from an unobserved axionlike behaviour of this particle,
they privilegiate a {\it \,pure vector\,} coupling of the $U$ boson to quarks and leptons, unless the corresponding symmetry is broken sufficiently above the electroweak scale.
\\[1mm]
{}
\end{abstract}

\vspace*{23mm}

\maketitle

\section{Introduction}

The $\,SU(3)\times SU(2)\times U(1)\,$ standard model gives a very good description of
strong and electroweak phenomena, so that the possible existence, next to the gluons, photon, $W^\pm$ and $\,Z$, of an additional neutral gauge boson,
called here the $U$ boson, is severely constrained.
The $U$ contributions to neutral-current amplitudes should be sufficiently small,
as well as its mixing with the $Z$.
According to the usual belief, any such new interaction
must be weaker than ordinary weak interactions,
or it would have been seen already.

\vspace{2mm}

This only applies directly, in fact, to heavy neutral gauge bosons.
For {\it \,light\,} gauge bosons
having {\it \,small\,}\, couplings to Standard Model particles, the discussion is different \cite{pfu}.
When the mass $\,m_U$ of the exchanged boson is small (compared to
the momentum transfer {\small $\sqrt{\,|q^2|}\,$}),
propagator effects are important.
$U$-induced cross sections then generally
{\it \,decrease with energy\,}, as for electromagnetic ones,
as soon as  $|q^2|$ gets larger than $\,\approx m_U^{\,2}$
(as for $Z$-exchanges, above the $Z$ mass),
and may be sufficiently small, if the $U$ couplings are small enough.
In particular, the existence of a new {\it light\,} gauge boson $U$ having couplings $f$
to matter particles such that
\be
\frac{f^2}{m_U^{\,2}}\ \ \sim\ \  \frac{g^2+g'^2}{m_Z^2}\ \ \
(\,\hbox{or}\ \sim \ G_F)\ \ ,
\ee
for example, is not necessarily excluded by {\it \,high-energy\,} scattering experiments.
Experiments performed at lower energies, such as
those measuring parity-violation effects in atomic physics
(as we shall discuss here),
or neutrino scattering cross sections at lower $|q^2|$,
are particularly relevant to search for
such a particle, and constrain its properties\,\cite{pfu,pfpv,bp}.

\vspace{2mm}

Let us recall, however, that even if it is {\it very\,} weakly coupled,
a light spin-1 $\,U$ boson could still have detectable interactions.
And this, even in the limit in which  its couplings $f$ to quarks and leptons would  almost vanish,
a very surprizing result indeed (apparently)\,!

\vspace{2mm}

In fact a very light spin-1 $U$ boson behaves in this case
($f \to$ very small, $\,m_U\to$ very small) very much as a quasimassless spin-0 axionlike particle,
if the current to which it is coupled includes a
(non-conserved) axial part\,\footnote{This is very similar to what happens in supersymmetry/supergravity theories, in which a very light spin-$\frac{3}{2}$ gravitino does not decouple in the $\,\kappa\to 0\,$ limit, but interacts (proportionately to $\kappa/m_{3/2}$ or $1/\Lambda_{\rm ss}^{\ \,2}$)
\,like the massless spin-$\frac{1}{2}$ goldstino of global supersymmetry
(a feature largely used in ``GMSB'' models)\,\cite{grav}.}.
This axionlike behavior then restricts rather strongly its
possible existence and properties,
implying that the corresponding gauge symmetry be broken at a scale at least somewhat above the electroweak scale; or even at a very high scale, according to the ``invisible $U$-boson'' mechanism\,\cite{pfu,pfu2}.

\vspace{1.5mm}

It is also possible that the new current to which the $U$ couples is {\it \,purely vectorial}, involving a linear combination of the conserved $B$, $L$ and electromagnetic currents, as in a class of models
discussed in Refs.~\cite{pfuV,pfu2}.
In this case there is no such axionlike behavior of the $U$ boson.
No significant extra contribution
to parity-violation effects is then to be expected.

\vspace{1.5mm}

We now turn to the recent suggestion of Light Dark Matter particles.
Contrasting with the heavy WIMPs, such as the neutralinos of supersymmetry,
{\it light\,} (annihilating) \,spin-$\frac{1}{2}$ or spin-0 particles can also
be acceptable Dark Matter candidates.
This requires, however, that they annihilate {\it very\,} efficiently, necessitating
new interactions, as induced by a light $\,U$ boson\,{\cite{bf,ldm}.
The required annihilation cross sections ($\,\approx$ 4 to 10 pb,
depending on whether Dark Matter particles are self-conjugate
or not),
must be significantly {\it \,larger\,} than weak-interaction cross sections
(for this energy), otherwise the relic abundance would be too large\,!
The $U$-induced Dark Matter annihilation cross section
into $e^+e^-$ ($\,\sigma_{ann} \,v_{rel}/c$)
\,also includes, naturally, a $v_{dm}^{\,2}$ low-energy suppression factor
(as desirable to avoid excessive $\gamma$ rays
from residual light Dark Matter annihilations  \cite{bes}).
This requirement is satisfied, in the case of a spin-$\frac{1}{2}$ Dark Matter particle
{\it axially\,} coupled to the $U$,
if this one is {\it \,vectorially\,} coupled to electrons\,\cite{ldm}.

\vspace{1.5mm}

A new interaction stronger than
weak interactions could seem,
naively, to be ruled out experimentally.
In fact, however, the $\,U$-mediated Dark-Matter/Matter interactions
should be {\it stronger than ordinary weak interactions} but only {\it at lower energies}, when weak interactions are really very weak.
But {\it weaker at higher energies}, \,at which they are damped by
$\,U$ propagator effects (for $s$ or $|q^2| > m_U^{\,2}$),
when weak-interaction cross sections,
still growing with energy like $s$, become important.
The smallness of the $\,U$ couplings to ordinary matter ($f$),
as compared to $\,e$, \,by several orders of magnitude,
and of the resulting $\,U$ amplitudes compared to electromagnetic ones,
\,can then account for the fact that these particles have not been observed yet.
The $U$ boson, in addition, may well have dominant invisible decay modes 
into unobserved Dark Matter particles.

\vspace{1.5mm}
We indicated in may 2003
that a gamma ray signature from the galactic centre at low energy could be
due to the existence of a light new gauge boson, inducing
annihilations of Light Dark Matter particles into $\,e^+e^-$ \cite{bf}.
The observation, a few months later, by the satellite INTEGRAL
of a bright \,511\, keV \,$\gamma$ ray line from the galactic bulge \cite{integral},
requiring a rather large number of annihilating positrons,
may then be viewed as originating from
Light Dark Matter annihilations~\cite{betal}.
Indeed \hbox{spin-0}, or as well spin-$\frac{1}{2}$ particles,
could be responsible for this bright \,511\, keV line,
which does not seem to have an easy interpretation in terms
of known astrophysical processes~\cite{ldm,bfsi}.
One should, however, also keep in mind that Light Dark Matter particles may still exist,
even if they are not responsible for this line.
(And that a light $U$ boson may be present, even if Light Dark Matter particles don't
exist at all\,\footnote{In addition, a $U$ that would be both extremely light
and extremely weakly coupled would lead to a new long-range force, and to the possibility
of (apparent) violations of the Equivalence Principe.}.)

\vspace{2mm}
Returning to Standard Model particles,
a new interaction that would be stronger than weak interactions at lower energies (at least when dealing with Light Dark Matter particle annihilations) could have important implications on ordinary physics,
{\it especially at lower energies or momentum transfer},
even if it has no significant influence on high-energy neutral current processes.

\vspace{2mm}

As we shall see,
parity-violation atomic-physics experiments\,\cite{pvexp} provide new strong constraints
on such a gauge boson \,-- whether light or heavy --\,
if its couplings to quarks and electrons violate parity,
then requiring that the corresponding
symmetry be broken significantly above the electroweak scale.

\section{\sloppy The effective weak charge \hbox{of a nucleus}}

Such models, in which the standard gauge group is extended to include
$\,SU(3)\times SU(2)\times U(1) \times \hbox{extra-}U(1)$ at least,
have been discussed in detail.
They involve an additional neutral gauge boson $U$ (which may also be called $Z'$ or $Z"$), initially associated,
before gauge symmetry breaking, with the extra-$U(1)$ generator. Mixing effects with the $Z$, however, in general play an important r\^ole, as far as the $U$ couplings are concerned\,\cite{pfu,pfuV,pfu2}. When the extra-$U(1)$ gauge coupling constant ($g"$)
is small compared to $g$ and $g'$, the mixing angle
({\small$\,\propto g"/$}{\footnotesize$\sqrt{g^2+g'^2}$}\,) turns out to be small
also,
and the modification to the $Z$ weak neutral current,
still given by $\,(J^\mu_{\ 3}-\sin^2\theta\,J^\mu_{\ em})\,$ up to very small corrections ({\small$\,\propto\, g"^2/(g^2+g'^2)$}), is in fact negligible.

\vspace{2mm}
The current to which the $U$ boson couples, however, is significantly
affected by the mixing, acquiring, in addition to the initial extra-$U(1)$
term, a new contribution proportional to $\,(J^\mu_{\ 3}-\sin^2\theta\,J^\mu_{\ em})\,$.
The resulting $U$-current
includes in general a {\it \,vector\,} part which appears as a linear combination
of the conserved $B$, $L$ and electromagnetic currents, as well as an {\it \,axial\,}
part (which may, however, not be present at all, depending on the models considered).
In particular, in a class of simple one-Higgs-doublet models
the quark-and-lepton contribution to the $U$ current
turns out to be {\it \,purely vectorial\,}\,\cite{pfuV,pfu2}
\,-- which also provides the desired $\,v_{dm}^{\,2}\,$ factor
in the annihilation cross section of spin-$\frac{1}{2}\,$ light Dark Matter particles
\cite{ldm}.

\vspace{2mm}

We now proceed with the phenomenological analysis,
expressing the relevant couplings in the Lagrangian density as follows:
\be
\ba{ccl}
\!\!{\mathcal L} &=&-\,e\ A_\mu\,J^\mu_{\ em}\,-\,Z_\mu\ \sqrt{\,g^2+g'^2}\ \left(\,J^\mu_{\ 3}\,-\,\sin^2\theta\,J^\mu_{\ em}\,\right)
\\[2.5mm]
&& \ \ \ -\ U_\mu\ \, \hbox{\large$\Sigma$}_{\,f=l,q}\ \ \bar f\,\gamma^\mu\,(f_{Vf}-\gamma_5\,f_{Af}\,)\,f \ .
\ea
\ee
The left- and right-handed projectors are
$\,{\mathcal P}_L = \frac{1-\gamma_5}{2}\,$, ${\mathcal P}_R = \frac{1+\gamma_5}{2}\,$
(so that a left-handed $\,U$-current would correspond to $\,f_V\!=f_A$),
\,with
$\ \gamma_5=$
\hbox{\tiny
$\left(\ba{cc} \hbox{\scriptsize 0}\! & \!\hbox{\scriptsize $ 1\!\!1$}\! \cr
\hbox{\scriptsize $ 1\!\!1 $}\! & \!\hbox{\scriptsize 0}\!  \ea \right)$}\,,
\normalsize
\,and the metric $( + - -\, -\, )\,$.
The relevant terms in the $Z$ weak neutral current are given by
\be
\ba{ccl}
J^\mu_{\ Z}&=&J^\mu_{\ 3}\,-\,\sin^2\theta\,J^\mu_{\ em} \vspace{3mm}\\
&=&
\ \ \frac{1}{4}\ \,\bar e\,\gamma^\mu\gamma_5\,e \,+\,
 (\,-\,\frac{1}{4}+s^2\,)\ \,\bar e\gamma^\mu e
\vspace{3mm} \\
&& \!\!\!-\ \frac{1}{4}\ \bar u\,\gamma^\mu\gamma_5\,u \, +\,
(\ \frac{1}{4}-\frac{2}{3}\,s^2\,)\ \,\bar u\gamma^\mu u
\vspace{3mm} \\
&& \!\!\! + \ \frac{1}{4}\ \bar d\,\gamma^\mu\gamma_5\,d \ +\,
(\,-\,\frac{1}{4}+\frac{1}{3}\,s^2\,)\ \,\bar d\,\gamma^\mu d\ \ ,
\\ [1mm]
\ea
\ee
with $\,s^2=\sin^2\theta=g'^2/(g^2+g'^2)$, $\,\theta\,$ being the electroweak mixing angle\,\footnote{We disregard here, as explained earlier, the very small
influence of $\,Z-U$ mixing effects on the $Z$ current.}.
The vector part of the $Z$ weak neutral current
is associated with the $Z$ (vectorial) weak charge, which reads, as far as the quark contribution is concerned\,\footnote{This may also be obtained from the quark vector couplings to the $Z$, as
$\
Q_Z=(2 Z+N)\,(\,\frac{1}{4}-\frac{2}{3}\,s^2\,)\,+\,(Z+2N)$  $(\,-\frac{1}{4}+\frac{1}{3}\,s^2\,)
\ \equiv\ \frac{1}{4}\ [\,Z\,(\,1-4\,s^2\,)\,-N\,]
\ =\ \frac{1}{4}\ \,Q_{W}(Z,\,N)_{SM}\,$.}
\footnote{More generally, $\,Q_Z=\frac{1}{2}\  T_{3\,(L+R)}-\,\sin^2\theta \ Q\,$ may be rewritten, using
$\, T_{3\,(L+R)}=\,Q\,-\,\frac{1}{2}\,(B-L)\,$, \,as the con\-served charge
\vspace{-2.5mm}\\
$$
Q_Z\,=\,-\ \frac{1}{4}\ (B-L) \,+\,(\ \frac{1}{2}\,-\,\sin^2\theta\,) \ Q\ \ ,
$$
\vspace{-2.5mm}\\
leading to a Standard Model ``weak charge''
$\ -\ (B-L)\,+\,(\,2\,-\,4\sin^2\theta) \ Q\,$,
\,identical to (\ref{qw}) in the case of a nucleus.
},
\be
\label{qz}
\ba{ccl}
\!\!\!\!Q_Z &=&
(T_{3\,L})_V-\sin^2\theta \ Q
\ =\
\hbox{\small$\displaystyle
\frac{T_{3\,L}+T_{3\,R}}{2}$}
\,-\,\sin^2\theta \ Q
\vspace{2mm} \\
&=&\
\hbox{\small$\displaystyle \frac{Z-N}{4}  $}
\,-\,\sin^2\theta \ Z\ \ =\ \ \frac{1}{4}\ Q_{\rm weak}\,(Z,\,N)\ \ .\\ [1mm]
\ea
\ee
The quantity
\be
\label{qw}
Q_W\,(Z,\,N)\ \,=\,\ Z\ (\,1-4\ \sin^2\theta\,)\,-\,N\ \ \,\approx\ \,-\,N\ \ ,
\ee
to which it is proportional,
is usually referred to as the ``weak charge'' of a nucleus of $Z$ protons and $N$
neutrons, and governs the parity-violation effects in atomic physics
we are interested in\ \cite{bou,flam}.

\vspace{2mm}

The corresponding effective Lagrangian density
involves the products of the ($Z$ and $U$) {\it \,axial\,} currents of the electron by the {\it vector\,} neutral currents of the quarks \,(i.e. ultimately the vector currents associated with
protons and neutrons). It may be written, in the local limit approximation
(assuming $m_U^{\,2}$  somewhat larger than the relevant $|q^2|$, \,cf.~\,Section \ref{sec:prop})
as:
\be
\label{leff}
\ba{ccl}
\!\!{-\ \mathcal L}_{\rm eff}&=&
\hbox{\footnotesize$\displaystyle
\frac{g^2+g'^2}{m_Z^{\,2}}  $}
\ \ \frac{1}{4}\ \,\bar e\,\gamma_\mu\gamma_5\,e \ \
\vspace{2mm}\\
&&
\ \ \ \ \left[ \,(\,\frac{1}{4}-\frac{2}{3}\,s^2\,)\ \,\bar u\gamma^\mu u\, +\,
(\,-\,\frac{1}{4}+\frac{1}{3}\,s^2\,)\ \,\bar d\,\gamma^\mu d\,
\right]
\vspace{4mm}
\\
&&-\ \,
\hbox{\footnotesize$\displaystyle
\frac{f_{Ae}}{m_U^{\,2}} $}
\ \ \bar e\,\gamma_\mu\gamma_5\,e\ \
\left[ \,f_{Vu}\ \bar u\gamma^\mu u\,+\,f_{Vd}\ \bar d\,\gamma^\mu d\,\right]
\ \,.
\ea
\ee

The quark (or proton and neutron) contribution to the charge $\,Q_U$ associated with the vector part of the $U$ current reads
\be
\label{effqcoup}
\!\!\!\!\!\ba{ccl}
\!\!\!\!\!Q_U&=&
\ (2f_{Vu}+f_{Vd})\ Z\,+\,(f_{Vu}+2f_{Vd})\ N
\vspace{3mm}\\
&=&\ 3\ f_{Vq}^{\rm \,eff}\ (Z+N)\ \ =\ \ 3\ f_{Vq}^{\rm \,eff}\ A\ \ .
\ea
\ee
This proportionality to the total number of nucleons $A$ holds only, strictly speaking,
when the $U$ has equal vector couplings to the $u$ and $d$ quarks, $f_{V\,u}=f_{V\,d}\,$.
If not, we can still use eq.\,(\ref{effqcoup}) as defining
the average effective vector coupling $\,f_{Vq}^{\rm \,eff}\,$ of the $U$ boson to a quark, within the nucleus considered.

\vspace{2mm}

The effective Lagrangian density (\ref{leff}) responsible
for atomic parity-violation effects leads to the parity-vio\-lating
Hamiltonian density for the electron field, in the vicinity of the  nucleus~\footnote{Finite-size effects of the nucleus may be taken into account
by replacing $\delta(\vec r\,)$ by the nuclear density $\,\rho_n(\vec r\,)$,
normalized to unity
(assuming here for simplicity that the $p$ and $n$ densities have the same radial behaviour).}:
\be
\label{heff}
\ba{ccl}
\!\!\!{\cal H}_{\rm eff}&=&e^\dagger(\vec r\,)\,\gamma_5\,e(\vec r\,)\
\hbox{\, \large $\left[\ \frac{g^2+g'^2}{16\,m_Z^{\,2}}\right.$}\ \ [\,Z\,(\,1-4\,s^2\,)\,-\,N\,]
\vspace{1.5mm} \\
&&\ \ \ \ \ \ \ \ \ \ \ \ \ \ \ \ \ \ \ \
\left. -\ \,\hbox{\large $\frac{f_{Ae}\,f_{Vq}^{\rm \tiny \ eff}}{m_U^{\,2}}$}\ \ 3\ (Z+N)
\ \right]\ \delta(\vec r\,)\!\!\!{}
\vspace{2mm} \\
&=&
\displaystyle
\ e^\dagger(\vec r\,)\,\gamma_5\,e(\vec r\,)\ \ \ \frac{G_F}{2\,\sqrt 2}\ \ \ Q_W^{\rm \,eff}(Z,\,N)\ \ \delta(\vec r\,)\ \ .
\ea\\[2mm]
\ee
In the non-relativistic limit (with small components expressed as $\,\simeq \vec \sigma.\,\vec p\,/(2\,m_e)$ acting on the electron wave-func\-tion), this turns into the parity-violating hamiltonian for an atomic electron,
\be
\label{heff2}
H_{\rm eff}\ \ =\ \
\displaystyle
\frac{G_F}{2\,\sqrt 2}\ \
\frac{\vec \sigma.\,\vec p\ \,\delta(\vec r)\,+\,\delta(\vec r)\,\ \vec \sigma.\,\vec p}{2\,m_e}
\ \ \ Q_W^{\rm \,eff}(Z,\,N)\ \ \ ,
\ee
$\,\vec p\,$ being the electron momentum operator.
\vspace{3mm}

This hamiltonian is expressed in terms of an ``effective weak-charge''
of the nucleus, which includes, in addition to the standard contribution
$\,Q_W(Z,\,N)_{SM}\,$
(given by eqs.\,(\ref{qz},\ref{qw}), plus radiative correction terms),
an additional \,$U$ contribution, in the case of a parity-violating $\,U$ current:
\be
\label{dqw0}
Q_W^{\rm eff}(Z,N)\,=\,
Q_W(Z,N)_{SM}\,-\,
\frac{2\sqrt 2}{G_F}\ \,\frac{f_{Ae}\,f_{Vq}^{\rm \,eff}}{m_U^{\,2}}\ \ 3\ (Z+N)\ .
\ee
This applies even if the vector coupling of the $U$
differs for the $u$ and $d$ quarks, the effective quark vector coupling $\,f_{Vq}^{\rm \,eff}\,$
being defined from eq.\,(\ref{effqcoup}) by
\be
f_{Vq}^{\rm \,eff}\ \ =\ \ \frac{f_{Vu}\ (2 Z+N)\,+\,f_{Vd}\ (Z+2N)}{3\,(Z+N)}\ \ \,.
\ee

\section{Expression in terms of the symmetry-breaking scale}

Equation (\ref{dqw0}), namely
\bea
\label{dqw}
\framebox [8cm]{\rule[-.5cm]{0cm}{1.3cm} $\displaystyle {
\Delta\,Q_W^{\rm eff}(Z,\,N)\ =\ -\ \frac{2\,\sqrt 2}{G_F}\ \,\frac{f_{Ae}\,f_{Vq}^{\rm \,eff}}{m_U^{\,2}}\ \ 3\, (Z+N)\ \ ,
}$}
\nonumber\\[-7mm]
\nonumber
\eea
\be
\ee
may be identified with the one given in \cite{pfpv},
\be
\label{dqw2}
\Delta\,Q_W\ \ =\ \ r^2\ \,c_\varphi\ \ 3\ (Z+N)\ \ ,
\ee
in a simple situation with
a universal axial contribution to the $U$ current.
The axial and vector couplings are then parametrized, in terms of the extra-$U(1)$ gauge coupling $g"$\,\footnote{The couplings to the left-handed and right-handed fermion fields were expressed as $\,-\frac{g"}{4}\,(1-c_\varphi),\ \frac{g"}{4}\,(1+c_\varphi)$,
respectively, which corresponds to a vector coupling $\,f_V=\frac{g"}{4}\,c_\varphi$,
and an axial coupling $\,f_A=-\frac{g"}{4}$.},
as
\be
\left\{\
\ba{ccccl}
\!\!-\,f_A &=& \hbox{\small $\displaystyle\frac{g"}{4}$} &=& \
2^{-\frac{3}{4}}\ \,G_F^{\ \frac{1}{2}}\ m_U \ r 
\vspace{2mm} \\ 
&&&&\ \ \ \simeq \ \,
2\ \,10^{-6} \ \,m_U(\hbox{\footnotesize MeV})\ \,r\ \ ,
\vspace{3mm}
\\
\ \ f_V  &=& \ \hbox{\small $\displaystyle \frac{g"}{4}$}\ \,c_\varphi &=& \
2^{-\frac{3}{4}}\ \,G_F^{ \ \frac{1}{2}}\ m_U \ r \ c_\varphi
\vspace{2mm} \\ 
&&&&\ \ \ \simeq \ \,
2\ \,10^{-6} \ \,m_U(\hbox{\footnotesize MeV})\ \,r\ c_\varphi\ \ .
\ea
\right.
\ee
$r\leq 1\,$ (here simply defined by $\,\frac{g"}{m_U}\,=\,\frac{g}{m_W}\ r$\,)
is a dimensionless parameter related to the extra-$U(1)$ symmetry-breaking scale, and
$c_\varphi\,$ (initially denoted $\,\cos\varphi\,$,
\,although not necessarily smaller than 1 in modulus)
measures the magnitude of the quark vector coupling relatively to the axial one.
The parity-violating $U$-exchange amplitudes are proportional to the
Fermilike constant
\be
-\ \frac{f_A\,f_V}{m_U^{\,2}}\ \ =\ \ \frac{g"^2}{16\,m_U^{\,2}}\ \,c_\varphi\ \ =\ \ \frac{G_F}{2\,\sqrt 2}\ \ r^2\ c_\varphi\ \ ,
\ee
allowing the identification of expressions (\ref{dqw}) and (\ref{dqw2}) of $\,\Delta Q_W$.

\vspace{2mm}
The parameter $r \leq 1$ represents more generally, in such models,
the scale at which the extra-$U(1)$ symmetry gets spontaneously broken (to which it is,
roughly, inversely proportional)\,\cite{pfu,pfu2}.
In a class of models $r=1$ would correspond to an extra $U(1)$ broken ``at the electroweak scale''
by two Higgs doublets only ($\,<\!\varphi_d^{\,\circ}\!>\ =v_1/\sqrt2\,$ and $\,<\!\varphi_u^{\,\circ}\!>\ =v_2/\sqrt2\,$);
this, however, is excluded experimentally as the light $\,U$ would then behave very much as a standard axion
(with presumably, in the present case, invisible decay modes into Light Dark Matter particles
dominating over the visible ones into $\,e^+e^-$).

\vspace{2mm}
$r\,$ is smaller than one, if an extra Higgs singlet
provides an additional contribution to the $U$ mass,
$\,r< 1$ measuring the amount by which the extra-$U(1)$ symmetry gets broken ``above the electroweak scale'' or ``at a large scale'' through this (large) extra singlet v.e.v.\,\footnote{In particular, if we define $\,v_2/v_1 =1/x= \tan\beta$, the axial coupling
$\,f_{A\,e}\,$ is given, after $\,Z-U\,$ mixing effects,
by $\,-\,f_{A\,e}= 2^{-\frac{3}{4}}\ G_F\,\! ^\frac{1}{2}\ m_U\ \frac{r}{x}$\,.}.
This can make the physical effects of the $U$ boson essentially invisible in particle physics, very much as for an axion, according to the ``invisible $\,U$ boson''
(or similar ``invisible axion'') mechanism\,\cite{pfu,pfu2}.

\vspace{2mm}
Reexpressing $\,\Delta\,Q_W\,$ as in (\ref{dqw2}) allows us to compare directly
the extra amount of parity-violation due to the $U$ boson
to the $Z$ contribution, in terms of $\,r\leq 1$.
The $U$ contribution
(for parity-violating couplings to quarks and electrons)
would be roughly of the same order as the standard one (and then
excessively large),
if the extra-$U(1)$ were broken at a scale comparable to the electroweak scale.

\section{Propagator effects for a \hbox{{\it\,very light\,}} $\,U$ boson}
\label{sec:prop}

\vspace{-1mm}

In addition, if the $\,U$ is light enough (i.e., as compared to the typical
{\small $\,\sqrt{\,|q^2|}\,$}
 in the experiment considered),
one can no longer use for its propagator the local limit approximation.
One writes instead,
\be
\label{prop}
\frac{f_{Ae}\,f_{Vq}^{\rm \,eff}}{m_U^{\,2}-q^2}\ \,=\,\
\frac{f_{Ae}\,f_{Vq}^{\rm \,eff}}{m_U^{\,2}}\ \ \frac{m_U^{\,2}}{m_U^{\,2}-q^2}\ \ ,
\ee
which leads to a corrective factor
\be
\label{cor}
\frac{m_U^{\,2}}{m_U^{\,2}-q^2}\ \,=\,\ \frac{m_U^{\,2}}{m_U^{\,2}+\vec q\,^2} \ \,\simeq\ \,
\hbox{\small $\left\{\,\
\ba{l} 0 \ \ \hbox{for}\ m_U^{\,2}\ \ll \ \vec q\,^2\,,\vspace{2mm} \\
1  \ \ \hbox{for}\ m_U^{\,2}\ \gg \ \vec q\,^2\,,\ea \right.
$}
\ee
as compared to a calculation that would have been performed
in the local limit approximation.

\vspace{2mm}

Expression (\ref{prop}) is associated with a Yukawa-like
(or Coulomb-like, if the $U$ is massless) parity-violating potential, i.e.
\be
\label{yuk}
\ba{c}
\displaystyle
\frac{f_{Ae}\,f_{Vq}^{\rm \,eff}}{m_U^{\,2}-q^2}\ \ \ \longleftrightarrow \hspace{5.8cm}
\vspace{1mm} \\ \ \
\displaystyle
f_{Ae}\,f_{Vq}^{\rm \,eff}\ \ \frac{e^{-\,m_U|\vec r\,|}}{4\,\pi\ |\vec r\,|}\ \,=\,\
\frac{f_{Ae}\,f_{Vq}^{\rm \,eff}}{m_U^{\,2}}\ \
\frac{m_U^{\,2}\ e^{-\,m_U|\vec r\,|}}{4\,\pi\ |\vec r\,|}\ \ ,
\ea
\ee

\vspace{-2mm}

\noindent
where
\be
\label{yuk2}
\frac{m_U^{\,2}\ e^{-\,m_U|\vec r\,|}}{4\,\pi\ |\vec r\,|}\ \ \
\left(
\hbox{\small $
\,=\,``\,\delta_{m_U}\!"\,(\vec r\,)\
$}
\right)\ \ \ \stackrel{\stackrel{\small\hbox{$m_U\to\infty$}}{}}{\longrightarrow}\ \ \ 
\delta\,(\vec r\,)\ \ ,
\ee

\vspace{-1mm}
\noindent
in the case of a sufficiently ``heavy'' $U$
(typically $m_U \simge 100$ MeV$/c^2$ as we shall see),
leading to the parity-violating hamiltonian (\ref{heff2})
(with $\,\delta\,(\vec r\,)$ replaced by the normalized nuclear density
$\,\rho_n(\vec r\,)$, if the nucleus is not taken as pointlike).

\vspace{2mm}
For a too light $U$ boson, however, the local limit approximation is not valid,
and the new contribution $\,\Delta\,Q_W^{\rm \,eff}\,$ \,as given by (\ref{dqw})
should be multiplied by a correction factor $\,K(m_U)$ obtained by replacing  $\,\delta(\vec r)$
in (\ref{heff}) or (\ref{heff2}) by the appropriate Yukawa distribution
$\,\delta_{m_U}(\vec r\,)$ of eq.\,(\ref{yuk2}),
which extends over a range $  \approx\hbar/(m_Uc)$.
The normalized nuclear density $\rho_n(\vec r_n)$ (which may be approximated by a $\delta(\vec r_n)$ distribution although it extends
over a radius $R_n\simeq r_0\,A^{1/3} \simeq 6 $ Fermi for Cs)
gets replaced by its convolution product with the $\,\delta_{m_U}(\vec r-\vec r_n)$\,
Yukawa distribution, corresponding to the exchange of a very light $U$ between an electron at $\,\vec r$ and the nucleus at $\vec r_n$.

\vspace{2mm}
This can be expressed through a corrective factor
\be
\ba{l}
\!\!K(m_U) \ =\
\displaystyle \frac{<\,{H_{\rm eff\ }}_{(m_U)}\,>}{<\,{H_{\rm eff\ }}_{(m_U\to\infty)}\,>}
\vspace{2mm}\\
\ \ \ \ \ \simeq\
\hbox{\footnotesize $\displaystyle
\frac
{{\int \,< \,e^{\dagger}(\vec{r}\,) \, \gamma_5  e (\vec{r})\,>\ \rho_n(\vec r_n\,)}\
\delta_{m_U}(\vec r-\vec r_n)\
d^3\vec r\ d^3\vec r_n}
{{\int\, < \,e^{\dagger}(\vec{r}\,) \, \gamma_5  e (\vec{r})\,>\ \rho_n(\vec r\,)}\ d^3\vec r}
$}
\ ,
\ea
\ee
evaluated in \cite{bp}, and given numerically in Table I.

The mass $\,m_U\simeq 2.4$ MeV$/c^2$, for which $K(m_U)=\frac{1}{2}\,$, defines the typical momentum transfer associated with cesium parity-violation experiments.
The corresponding $\,\hbar/(m_U\,c)$ is $\,\simeq$ 80\, Fermi:
the electron involved in the parity-violating transition of the cesium atom
``feels'' in fact the new $U$-mediated interaction, even if it is relatively long-ranged, essentially in the vicinity of the nucleus, where the screening of the Coulomb potential of the nucleus by the core electrons can be neglected.
One has therefore, ultimately,
\bea
\label{dqw3}
\framebox [8.6cm]{\rule[-.5cm]{0cm}{1.2cm} $\displaystyle {
\Delta\,Q_W^{\rm eff}(Z,\,N)\ =\,-\,\frac{2\,\sqrt 2}{G_F}\ \frac{f_{Ae}\,f_{Vq}^{\rm \,eff}}{m_U^{\,2}}\ \,3\, (Z+N)\ K(m_U)\,.
}$}
\nonumber\\[-6mm]
\nonumber
\eea
\be
\ee
For $\,m_U< 100$ MeV$/c^2$ \,the presence of the factor $K$  weakens the expressions
of the limits that would otherwise be obtained from (\ref{dqw}),
especially in the case of a very light gauge boson\,\footnote{Furthermore in
the limit of a {\it very light\,} $U$ (i.e. for
$m_U\ll \,m_e\,\alpha\,\simeq \,4\,$ keV$/c^2$ so that
$\,\hbar/(m_U\,c) \gg \hbar/(m_e\,c\,\alpha)$
$\simeq .5\ 10^{-8}$ cm),
$\,K(m_U)\propto m_U^{\,2}\,$ (as one can see from (\ref{cor})), and the limits
may then be expressed as $\,|f_{Ae}\,f_{Vq}^{\rm \,eff}\,|< \,...\ $,
\,instead of $\,|f_{Ae}\,f_{Vq}^{\rm \,eff}\,|/m_U^{\,2}$ $< \,...\ $.}.
Still they remain of the same order as obtained from a local limit approximation,
for $m_U\simge\,$ a few MeV$/c^2$\,'s, as can be seen from Table I.

\begin{table}[t]
\caption{Atomic factor $K(M_U)$ giving the correction to the weak charge  $ \Delta Q_W (m_U)$ for the cesium atom.}
\begin{center}
\footnotesize
\begin{tabular}{|c||c|c|c|c|c|c|c|c|c|c|}
\hline  &&&&&&&&&& \\ [-2.5mm]
$\ba{c} m_U \\ (\,\hbox{MeV}/c^2\,)\ea  $ &
\ .1 \  &\  .37 \  & \ .5 \ &\  \,1\, \  & \,2.4\, &\  \,5\, \  &\  10 \ & \ 20 \ &
\ 50 \  & \,100\, \\ [4mm]
\hline
&&&&&&&&&& \\ [-2mm]
$ \,\ba{c}\hbox{corr.~factor}\vspace{.5mm}\\ \ K(m_U)  \ea\, $
&\,.025\,  & \,.15\, & \,.20\, & \,.33\, & .5 &\,.63\,&\,.74\,&\,.83\,&\,.93\,&\,.98\,\\ [-2mm]
 &&&&&&&&&& \\
\hline
\end{tabular}
\vspace*{-2mm}
\end{center}
\end{table}

\section{Experimental limits on $\,f_{Ae}\,f_{Vq}\,$}

From the present comparison between experimental measurements of $\,Q_W(Z,\,N)$
for cesium, and theoretical predictions from standard model estimates\,\cite{pvexp,flam},
\be
\left\{
\ba{ccl}
Q_{W\,\rm exp}&=& -\ 72.74\ (29)_{\rm exp}\,(36)_{\rm theor}\ \ ,
\vspace{2mm} \\
Q_{W\,SM}&=&\ -\ 73.19\,\pm\,.13\ \ ,
\ea \right.
\ee
one gets
\be
\Delta\,Q_W\ \ =\ \  Q_{W \,\hbox{\small exp}}\,-\,Q_{W\,SM}   \ \ =\ \ 0.45\,\pm 0.48\ \ ,
\ee
which corresponds to an uncertainty of less than 1\,\%,
\,i.e. (working conservatively at a pseudo ``$\,2\,\sigma\,$'' level)
\be
\label{limdqw}
-\ .51\ < \ \Delta\,Q_W\ \ <\ 1.41\ \ .
\ee
For cesium ($A=133$), using
$\,G_F /(2\,\sqrt 2 \ \,3 A)\, \simeq\, 1.03$ $10^{-14}$ MeV$^{-2}\,
$
and expression (\ref{dqw}) of $\,\Delta\,Q_W$
(for $\,m_U$ $\,\simge\,$ 100 MeV$/c^2$), we get the constraint
\be
-\ .53\ 10^{-14}\ \ \hbox{MeV}^{-2}\ <\ \frac{-\,f_{Ae}\,f_{Vq}}{m_U^{\,2}}\ <\ 1.46\ 10^{-14}\ \ \hbox{MeV}^{-2}\ \ ,
\ee
or, approximately,
\be
\label{pvcont}
\framebox [8cm]{\rule[-.45cm]{0cm}{1.1cm} $\displaystyle
-\ .5\ 10^{-3}\ G_F\ <\ \frac{-\,f_{Ae}\,f_{Vq}}{m_U^{\,2}}\ <\ 1.3\ 10^{-3}\ G_F\ \ .
$}
\ee
For a lighter $U$ the limits get divided by the corrective factor $\,K(m_U)\,$
of Table I (i.e.~approximately doubled, for $m_U \simeq \,$ a few MeV$/c^2$\,'s).

\vspace{3mm}
This analysis applies to heavy as well as to light $U$'s.
In the first case it can constrain a $\,U$ (unmixed with the $Z$,
with couplings $\,\sim g,\ g'$ or $e$) to be heavier than several hundred GeV$/c^2$\,'s
or even more, depending on its couplings.
As a toy-model illustration, an extra $Z$ or $U$ boson that would have the same couplings as the $Z$  would lead directly to a negative contribution
$\,\Delta Q_W $ $\simeq Q_{W\,SM}\ (m_Z/m_{{\,\rm extra} \ Z})^2\,$.
Assuming for simplicity that no other contribution has to be considered,
it would have to verify, approximately, $\,|\Delta Q_W|<.55\,$.
The new gauge boson should then be at least $\,11.5$ times heavier than the $Z$, i.e.:
\be
m_{{\,\rm extra} \ Z}\ \,>\ \,1.05\ \,\hbox{TeV}/c^2\ \ ,
\ee
which is above present direct collider bounds.

\vspace{3mm}
For a light $U$ on the other hand, the constraint (cf. (\ref{pvcont}))
adds to those already obtained
from low-energy \hbox{$\,\nu-e\,$} scattering cross sections, e.g.,
for $m_U$ larger than a few MeV$/c^2$\,'s,
\be
\hbox{\small $\displaystyle \frac{|f_{V\nu}\,f_{Ve}|}{m_U^{\,2}}$}
\ \simle\ G_F\ \ ,
\ee
and anomalous magnetic moments of charged leptons\,\cite{pfu,pfpv,bf,ldm}.
The latter constraints, however, should be considered with appropriate care,
especially in the case of parity-violating couplings,
due to the possibility of cancellations between (positive) vector contributions and
(negative) axial ones \footnote{While the $g-2$ constraints on the vector and axial couplings to the electron, and vector coupling to the muon, are in general not so restrictive
(e.g. for a $U$ somewhat heavier than $e$ but lighter than $\mu$, $\,f_{V\,e}\simle 2\ 10^{-4}\  m_U$\,(MeV),
$\,f_{A\,e}\simle .6\ 10^{-4}\ m_U$\,(MeV),
$\,f_{V\,\mu} \simle\,6\ 10^{-4}$),
the one for an axial coupling to the muons
\,($f_{A\,\mu}\simle \,3\ 10^{-6}\ m_U$\,(MeV) \
i.e. $\,f_{A\,\mu}^{\ 2}/m_U^{\,2}\,<\,G_F$)\,
is more severe, in connection with an axionlike behavior of the $U$ boson in this case.}
\footnote{A light $U$ could also be detected through a brems\-strahlung from an electron,
in electron beam dump experiments, as for an axion decaying into $e^+e^-$
(but with a production cross section behaving differently, for a $U$ having vector or pseudovector couplings); 
this may constrain the $U$ to be heavier than $\,\sim$ a few to 10 MeV, depending on 
the size of its couplings\,\cite{dav}. However, a relatively light $U$ responsible for Light Dark Matter annihilations at the appropriate rate tends to be much more strongly coupled to Dark Matter ($c_U$) than to ordinary matter ($f$), possibly by several orders of magnitude.
Invisible $U$ decays into Dark Matter particles would both decrease significantly 
the $U$ lifetime and 
make its visible decays into $\,e^+e^-$ very rare, or even practically negligible;
no such limits may then be obtained in this way.}.

\vspace{2mm}

More significant in fact are the limits from the non-observation of an axionlike particle, which severely constrain an axial contribution in the quark $\,U$ current, requiring typically
\be
\label{limaq}
\frac{f_{A\,q}^{\ 2}}{m_U^{\,2}}\ <\ \frac{1}{10}\ \,G_F\ \ ,
\ee
from $\ \psi\,$ or $\,\Upsilon \,\to\,\gamma +U\,$ decays,
or even
\be
\frac{f_{A\,s}^{\ 2}}{m_U^{\,2}}\ \simle\ \frac{1}{300}\ \ G_F\ \ ,
\ee
from $\,K^+\to\,\pi^+U$ decays~\cite{pfu,ldm}.
If such an axial contribution is actually present, the extra-$U(1)$ symmetry
should then be broken  sufficiently above the electroweak scale \,--
a conclusion reenforced here, in the case of a $U$ boson
inducing atomic-physics parity-violation effects, constrained to be
very small.
This illustrates, also, how parity-violation atomic physics experiments
can give very valuable informations, complementing those obtained from
particle physics.

\section{Conclusion}

This analysis of parity-violation effects in atomic physics
(which also applies to heavy bosons), combined, in the case of a light $U$, with earlier constraints on a possible axionlike behavior of this particle,
favors a situation in which the quark-and-lepton contribution to the $U$ current
is {\it purely vectorial\,}, as in a class of models discussed in \cite{pfu2,pfuV}.
Otherwise the scale at which the extra-$U(1)$ symmetry is broken should be larger than the electroweak scale, by about one order of magnitude at least;
the coupling of the $U$ to a Light Dark Matter particle would then have to be further increased, to compensate for its smaller couplings to ordinary particles.


\begin{thebibliography}{99}



\bibitem{pfu} P. Fayet, \PLB 95, 285 (1980); \NPB 187, 184 (1981).

\bibitem{pfpv} P. Fayet, \PLB 96, 83 (1980).

\bibitem{bp} C. Bouchiat and C.-A. Piketty,  \PLB  128, 73 (1983).

\bibitem{grav} P. Fayet, \PLB 70, 461 (1977); B86, 272 (1979).

\bibitem{pfu2} P. Fayet, \NPB 347, 743 (1990); Class. Quant. Grav. 13, A19 (1996).

\bibitem{pfuV} P. Fayet, \PLB 227, 127 (1989).

\bibitem{bf} C. Bo$\!e$hm and P. Fayet, \NPB {683}, 219 (2004)
(hep-ph/0305261).

\bibitem{ldm} P. Fayet, \PRD 70, 023514 (2004) (hep-ph/0403226); hep-ph/0408357.

\bibitem{bes} C. Bo$\!e$hm, T. Ensslin and J. Silk, J. Phys. G 30, 279 (2004) (astro-ph/0208458).

\bibitem{integral} J. Kn\"odlseder \textit{et al.}, astro-ph/0309442;
P. Jean \textit {et al.}, Astron. Astrophys. 407 (2003) L55.

\bibitem{betal} C. Bo$\!e$hm {\textit {et al.}},
{\em Phys. Rev. Lett.} {92}, 101301 (2004) (astro-ph/0309686).

\bibitem{bfsi} C. Bo$\!e$hm, P. Fayet and J. Silk, \PRD 69, 101302 (2004) (hep-ph/0311143);
M. Cass\'e \textit{et al.}, Proc. 5th INTEGRAL Workshop, astro-ph/0404490,
ESA-SP-552; J. Beacom, N. Bell and G. Bertone, astro-ph/0409403.

\bibitem{pvexp} C. S. Wood \textit{et al.}, Science 275, 1759 (1997).

\bibitem{bou} M.-A. Bouchiat and C. Bouchiat, \PLB 48, 111 (1974);
J. de Physique 34, 899 (1974).

\bibitem{flam} J.S.M. Ginges and V.V. Flambaum, Phys. Rep. 397, 63 (2004).

\bibitem{dav} M. Davier, J. Jeanjean and  H. Nguyen Ngoc, \PLB 180, 295 (1986);
E. Riordan et al., \PRL 59, 755 (1987); M. Davier, XXIst
Conf. on Neutrino Physics and Astrophysics, Paris (june 2004),
http://neutrino2004.in2p3.fr/slides/friday/davier.pdf.



\end{thebibliography}
\end{document}